\documentclass[twocolumn,showpacs,preprintnumbers,amsmath,amssymb]{revtex4}
\usepackage{graphicx}% Include figure files
\usepackage{dcolumn}% Align table columns on decimal point
\usepackage{bm}% bold math

\def\bea{\begin{eqnarray}}
\def\eea{\end{eqnarray}}

\begin{document} 

\preprint{Version 1.6}

\title{ZYAM and $\bf v_2$: Underestimating jet yields from dihadron azimuth correlations}

\author{Thomas A. Trainor}
\address{CENPA 354290, University of Washington, Seattle, WA 98195}

%%%%%%%%%%%%%%%%%%%%%%%%%%%%%%%

\date{\today}

\begin{abstract}
Dihadron azimuth correlations can provide combinatoric access to jet structure in nuclear collisions. To isolate true jet yields a background must be subtracted, including a constant offset and a contribution from ``elliptic flow'' (azimuth quadrupole measured by $v^2_2$). The principle of ``zero yield at minimum'' (ZYAM) has been introduced to determine the constant offset. Independent measurements determine $v_2^2$. This analysis demonstrates that the ZYAM concept is invalid (offset typically overestimated) and $v_2^2$ is also overestimated by conventional measurements. Jet yields are thus substantially underestimated in more-central A-A collisions, and the ``away-side'' azimuth peak (back-to-back jet correlations) is strongly distorted, leading to incorrect inference of ``Mach shocks.''
\end{abstract}


\maketitle

%%%%%%%%%
 \section{Introduction}

The nominal goal of the Relativistic Heavy Ion Collider (RHIC) is production of quark-gluon plasma (QGP), a color-deconfined, thermalized state of quarks and gluons~\cite{qgp}. Several characteristic indicators for QGP formation have been proposed, one being modification of parton scattering and fragmentation to hadron jets in more-central A-A collisions~\cite{quench}. Colored partons should be uniquely sensitive to the presence and properties of a color-deconfined medium. In A-A collisions jets are studied combinatorially through {\em dihadron correlations} analyzed and interpreted in the context of parton energy loss in a dense medium~\cite{dihadron}. Two diametrically-opposed pictures have emerged from jet correlation analysis.

In the conventional picture the scattered-parton spectrum extends down as far as 1 GeV, resulting in copious parton production near that energy (estimating $\sim$1000 gluons per unit rapidity)~\cite{cgc,cooper}. Most partons deposit a large fraction of their energy in a ``medium'' and are thermalized~\cite{cooper,nayak}, producing a large energy density with accompanying pressure gradients which in turn drive large flow magnitudes (longitudinal, radial, elliptic)~\cite{finns}. 

The medium forms an ``opaque core'' which can stop energetic partons and which is surrounded by a less-dense ``corona'' from which single partons may escape radially outward while their in-going partners are stopped in the medium~\cite{core1,core2}. That picture is said to account for systematic modification of {\em triggered} dihadron correlations (usually with asymmetric $p_t$ cuts)~\cite{staras} and single-particle spectra (jet quenching and high-$p_t$ suppression)~\cite{starsupp}.

However, recent analysis conflicts with the existence of a thermalized dense partonic medium and presents a different picture of A-A collisions. {\em Untriggered} angular correlations (no $p_t$ cuts applied)~\cite{axialci,porter,daugherity,ptscale,ptedep} reveal that far from being ``quenched'' or lost to thermalization essentially all initially-scattered partons survive to produce correlated hadrons, albeit jets are significantly modified in more-central A-A collisions. Jet correlations are dominated by {\em minijets}~\cite{minijet}. Strong back-to-back (mini)jet correlations are still observed for all collision systems~\cite{daugherity}. Two-component analysis of $p_t$ spectra reveals that the hard component is quantitatively described by pQCD for all A-A centralities, and the full parton spectrum is manifested as correlated fragments in the final state~\cite{2comp,fragevo}. The same hadron spectrum analysis also indicates that the scattered parton spectrum terminates near 3 GeV, not 1 GeV, implying a factor 25 less parton production (less density) than the conventional picture.

Analysis of dihadron correlations typically incorporates several important assumptions: 1) background subtraction is determined in part by {\em zero yield at minimum} (ZYAM)~\cite{zyam}. 2) A contribution from ``elliptic flow'' is determined by independent measurement of an azimuth quadrupole component in the form $v_2^2$~\cite{v2}. 3) $p_t$ cuts define correlations from ``hard scattering'' as opposed to soft processes which could contaminate jet correlations. Certain features of dihadron correlations and interpretations which support parton energy loss in a dense, thermalized medium depend critically on those assumptions.

It is important to resolve the striking incompatibility between two collision models emerging from the same underlying RHIC data. In this analysis I re-examine dihadron correlation analysis methods, with special attention to the ZYAM assumption, $v_2$ subtraction and the consequences of trigger/associated $p_t$ cuts.

%%%%%%%%%%%%%%%
\section{Dihadron Azimuth correlations}

Modification of parton scattering and jet formation in more-central A-A collisions is seen as a possible signal of QGP formation. Jets are observed as {\em per-trigger} azimuth correlations by combinatoric reconstruction. Background subtraction, including ZYAM and $v_2$ estimation, plays a central role in jet correlation analysis and interpretation.

\subsection{Dihadron correlation analysis}

High-$p_t$ dihadron azimuth correlations are used to study possible jet modifications in \mbox{A-A} relative to \mbox{p-p} collisions. The density of particle pairs which satisfy certain $p_t$ (transverse momentum) cuts (``trigger'' and ``associated'' particles) is plotted on azimuth difference $\Delta \phi = \phi_\text{associated} - \phi_\text{trigger}$. Trigger particles fall within one of two (usually asymmetric) $p_t$ intervals. Combinatoric jet reconstruction is a proxy for true event-wise jet reconstruction in A-A collisions with large multiplicities.

A combinatoric background must be subtracted from the ``sibling'' (same-event) pair density to obtain the true distribution of jet-correlated pairs. The background is assumed to be modulated by an elliptic flow ($v_2$) contribution. $v_2$ is estimated by separate $v_2$-specific analysis. The zero reference is determined conventionally by the ZYAM assumption. In p-p and peripheral A-A collisions the dominant structures are a same-side (SS, $|\Delta \phi| < \pi / 2$) peak (nominally Gaussian), an away-side (AS, $|\Delta \phi -\pi| < \pi / 2$) peak, and a $\cos(2\Delta \phi)$ sinusoid measured by $v^2_2$ (azimuth quadrupole).

The ZYAM subtraction applied to more-central A-A collisions produces apparently interesting structures in the away-side component (nominally back-to-back jets) interpreted by some in terms of ``Mach cones,'' possibly the result of energetic partons interacting with a dense QCD medium~\cite{mach,machth1,machth2}. However, the supporting arguments for ZYAM are questionable, and values of $v_2$ used in the subtraction may be substantially overestimated leading to distortions of true jet correlations.

\subsection{Jet phenomenology}

An initial analysis of jet-related azimuth correlations in 200 GeV 0-10\% central Au-Au collisions compared to \mbox{p-p} collisions imposed a high-$p_t$ cutoff (2-4 GeV/c) on the associated-particle interval as a means to insure that true jet phenomena dominated the correlations~\cite{staras}. Combinatoric background subtraction invoked ZYAM and utilized values of $v_2(p_t)$ from independent measurements. Mean value $v_2 = 0.07$ was used for the subtraction.  The correlation analysis followed observation of high-$p_t$ suppression in single-particle spectra~\cite{starsupp}. The major conclusion was ``disappearance of the away-side jet'' (apparent back-to-back jet suppression).

Followup studies~\cite{starprl,phenix} sought to determine where the energy ``lost'' at larger $p_t$ appears in the final state. Jet correlations below 4 GeV/c were said to be affected by hydrodynamic flows and quark coalescence believed to dominate hadron production in that $p_t$ interval~\cite{phenix}. Combinatoric techniques permitted extension of  associated-particle $p_t$ cuts to as low as 0.15 GeV/c.

Azimuth correlations from conventional (ZYAM) analysis have the following characteristics: When smaller-$p_t$ associated particles are included both SS and AS peaks are broadened on azimuth in more-central Au-Au collisions, and the AS $p_t$ spectrum is ``softened''~\cite{starprl}. The SS peak is enhanced relative to p-p, and the AS peak develops a minimum (dip) at $\Delta \phi = \pi$. Those systematics are thought to reveal ``characteristics of the energy transport of the quenched partons'' on both $p_t$ and $\Delta \phi$~\cite{phenix}.

\subsection{Conventional physics interpretation}

The results of~\cite{staras} were interpreted to reveal an opaque ``core'' (dense medium) formed in more-central Au-Au collisions which stops most partons produced in or passing through that region. The core was interpreted as a thermalized QCD medium with large matter and energy densities~\cite{core1,core2}. Partons which interact strongly in the medium lose most of their energy by radiating gluons and are thermalized~\cite{phenix}. ``The {\em away side jet} traverses a large amount of matter'' [emphasis added]~\cite{starprl}. ``Shower'' gluons may be emitted at large angles, possibly resulting in novel jet structures~\cite{machth1,machth2}.  pQCD calculations of parton energy loss require 30 times normal nuclear gluon density to account for central Au-Au results~\cite{vitev}.

In the {\em core-corona} model of A-A collisions some partons are produced in a less-dense ``corona'' region outside the dense core and may survive to fragment outside the A-A system~\cite{phenix}. High-$p_t$ hadrons and hadron pairs mainly come from such partons, which suffer minimal interaction with the medium. Partons with $p_t > 5$ GeV/c are said to exhibit in-vacuum fragmentation based on 1) a p-p-like single-particle power-law spectrum and/or suppression (e.g., $R_{AA}$~\cite{raa}) {\em independent of} $p_t$, and 2) p-p-like azimuth correlations with distinct SS and AS Gaussians.

According to~\cite{starprl} increased soft-particle production in Au-Au compared to p-p indicates parton thermalization. Final hadron remnants  in central Au-Au collisions no longer exhibit jet-like correlations.  Persistence of a broad AS peak structure at smaller $p_t$ indicates global momentum conservation. Interactions seem to drive particles from jet fragmentation and from the bulk medium toward mutual equilibrium, possibly implying a high degree of thermalization of the ``medium.'' A significant amount of associated-particle energy may come {\em from the medium} through recombination, scattering or flow. 

\subsection{Alternative interpretations}

Recent analysis suggests alternative interpretations of spectra and angular correlations. Two-component analysis of 200 GeV Au-Au spectra~\cite{2comp,fragevo} suggests that the spectrum hard component represents fragments from the entire initial-state parton spectrum. Fragmentation functions are modified in more-central A-A collisions, but {\em no partons are lost to thermalization}. Analysis of 2D angular correlations~\cite{daugherity} is consistent with that picture: minijet correlation strength is proportional to N-N binary collisions ($n_{binary}$) for peripheral collisions, but {\em increases more rapidly than $n_{binary}$} for more-central collisions.

Triggered azimuth correlations are not easily compared with {\em normalized per-participant} minimum-bias results. Such comparisons could serve to test the extent of parton thermalization by direct comparison to pQCD predictions~\cite{fragevo}. 
High-$p_t$ trigger particles are themselves subject to changes in A-A dynamics with centrality. In contrast, absolute spectrum and correlation measures reveal that while AS correlations are strongly altered, scattered partons are not thermalized (stopped in a medium). 

The SS jet width in 1D azimuth correlations does not increase significantly with A-A centrality as would be expected from strong parton interactions with a dense medium (multiple scattering or gluon bremsstrahlung). In fact, for minimum-bias 2D correlations the SS peak azimuth width is observed to {\em decrease} with increasing centrality~\cite{daugherity}. No SS jet structure corresponds to ``Mach cones.'' Persistence of a broad AS peak at  smaller $p_t$ has been attributed to global momentum conservation~\cite{starprl}, but the AS peak amplitude scales with binary collisions, not with participant pairs as would be expected for global conservation.

Imposition of a 2-4 GeV/c associated cut in~\cite{staras} led to the conclusion that the AS jet is eliminated in more-central A-A collisions, implying absorption (thermalization) of one of the scattered partons. However, rejecting associated particles in the 0.15-2 GeV/c interval introduces a large bias. Extension of the AS $p_t$ interval down to 0.15 GeV/c reveals copious AS jet correlations. AS correlation structure is dominated by {\em inter\,}jet or jet-jet (parton-parton) correlations, not {\em intra\,}jet correlations. AS peak modifications per se may not reflect changes in parton abundance or fragmentation.

Effects of $k_t$ broadening may dominate AS correlation structure. The role of initial-state parton multiple scattering in A-A collisions is still not fully appreciated, although $k_t$ broadening has been studied with Monte Carlo simulations~\cite{staras}. The role of independently-measured $v_2$ in combinatoric background subtraction which isolates jet correlations is also not well-defined. Oversubtraction of $v_2$ can (and does) result in substantial artifacts. 

Production of a dense QCD medium could indeed result in jet quenching and other QGP ``signals,'' but modifications to parton scattering and fragmentation observed to date don't {\em require} production of a QGP. Given the importance of angular correlations and jet-related phenomena to interpretation of RHIC data we should reconsider the conventional picture in light of recent results.

%%%%%%%%%
 \section{The ZYAM Subtraction Procedure}

The main object of ``triggered'' dihadron correlation analysis is systematic study of jet structure variations with A-A centrality and trigger/associated combinations of $p_t$ cuts. Accurate isolation of jet correlations from other correlation structure is technically challenging. The distribution of ``sibling'' pairs (pairs within single events) averaged over an event ensemble contains contributions from several sources: 1) uncorrelated combinatoric background, 2) back-to-back jet pairs, 3) isolated jets (partner outside the $\eta$ acceptance), 4) an azimuth quadrupole component conventionally attributed to elliptic flow and 5) small contributions from other sources (e.g., quantum correlations, electron pairs from gamma conversions). 

To isolate jet contributions 2) + 3) in the ZYAM procedure a subtraction is performed. A nominally uncorrelated mixed-pair reference distribution 1) approximates the combinatoric background structure (e.g., acceptance/efficiency azimuth variations). But normalization of mixed pairs relative to sibling pairs depends in part on fluctuation sources (e.g., impact parameter fluctuations, large-scale correlations) which are not reliably estimated at present. The normalization is instead determined by the ZYAM criterion. The quadrupole component 4) is estimated from measurements of $v_2(p_t)$ by a variety of methods which are susceptible to various ``nonflow'' errors (mainly jet correlations). Background 1) is then modulated by the estimated quadrupole amplitude, assuming a thermalized flowing bulk medium.

 \subsection{Dihadron correlations}

Absent particle identification two-(charged)-{\em particle} correlations are measured. Electron pairs due to photon conversions from $\pi^0$ decays appear near the angular origin, are prominent in 2D angular autocorrelations~\cite{daugherity} and may distort the 1D same-side (jet) peak on azimuth. The electron-pair structure is narrow on pseudorapidity as well as azimuth difference, and is thus suppressed to some extent in the 1D azimuth projection.

Dihadron correlations have three main components: 1) a same-side (SS) peak centered at the origin, nominally Gaussian in shape and representing {\em intra}\,jet correlations, 2) an away-side (AS) peak centered at $\pi$ and representing {\em inter}\,jet correlations from back-to-back jets, 3) a $\cos(2\Delta \phi)$ term conventionally measured by $v_2^2$. The $\cos(2\Delta \phi)$ term is the {\em azimuth quadrupole} component~\cite{newflow}.

\subsection{Trigger/associated $p_t$ cuts}

``Triggered'' jet correlation analysis emulates event-wise jet reconstruction in the high-multiplicity environment of A-A collisions by imposing $p_t$ cuts on particle pairs. A trigger particle at larger $p_t$ is intended to estimate the leading-parton momentum. An associated particle at smaller $p_t$ should then sample the jet fragment distribution. Aside from that model-dependent language the resulting azimuth correlations correspond to specific cut boxes on the space $(p_{t1},p_{t2})$. Minimum-bias or ``untriggered'' jet correlations are obtained by applying no $p_t$ cuts other than an acceptance cut near 0.1 GeV/c.

An asymmetric cut system has some validity for trigger $p_t > 4$ GeV/c where the jet multiplicity is significantly larger than 2. For the majority of jets (minijets) that condition is not satisfied. A symmetric $p_t$ cut then minimizes bias~\cite{porter}. Asymmetric $p_t$ cuts imply significant distance from the diagonal of the $(p_{t1},p_{t2})$ cut space, which has implications for joint $v_2^2(p_{t1},p_{t2})$ distributions compared to marginal distributions and factorization assumptions.

 \subsection{$\bf v_2$ estimation}

Azimuth correlation measure $v_2$ is estimated by several methods denoted by $v_2$\{\text{method}\}, e.g. $v_2\{2\}$ (two-particle correlations), $v_2\{EP\}$ (event plane or ``standard'' method), $v_2\{4\}$ (four-particle cumulants), etc.~\cite{flowmeth}. In practice, different methods return different $v_2$ values for the same physical conditions, depending on the amount of ``nonflow'' (minijets) confused by each method with the intended quadrupole component~\cite{newflow,gluequad}. 

Various strategies are employed to reduce contributions from nonflow and $v_2$ fluctuations. $v_2\{4\}$ is said to eliminate nonflow contributions assuming nonflow is ``clusters'' of 2-3 particles (inconsistent with minijet systematics in more-central Au-Au collisions). Average $(v^2_2\{2\} + v^2_2\{4\})/2$ is said to eliminate contributions from $v_2$ fluctuations based on several physical assumptions~\cite{gluequad}. In the simulations presented here $v^2_2\{2\}$ is determined by a fit (light dotted curve) to the entire azimuth distribution, and the background quadrupole amplitude $v_2^2$ is defined by  $v_2^2\{2\}/2$ (assuming $v^2_2\{4\} = 0$ for central A-A collisions). $v_2^2$ is in that case locked to the amplitude of the SS Gaussian.

The quadrupole contribution relevant to two-particle correlations subject to $p_t$ cuts is determined by joint distribution $v_2^2(p_{t1},p_{t2})$, which has not been measured. In the conventional approach factorization is assumed---$v_2^2(p_{t1},p_{t2}) \approx v_2(p_{t1}) \, v_2(p_{t2})$, where $v_2(p_t)$ is from a separately-measured marginal distribution.

\subsection{ZYAM background subtraction}

The per-event and per-trigger ``raw'' pair density is
\bea \label{raw}
R(\Delta \phi ;p_{t1},p_{t2}) \hspace{-.03in} &=& \hspace{-.03in} \left\langle \frac{1}{ N_{trig}} \sum_{i \in p_{t1},j \in p_{t2}} \hspace{-.2in} \delta(\phi_i - \phi_j -\Delta \phi)\right \rangle
\eea
averaged over an event ensemble and integrated over pseudorapidity acceptance $\Delta \eta$, where $p_{t1}$ and $p_{t2}$ denote ``trigger'' and ``associated'' cut bins and $N_{trig} \equiv \sum_{i \in p_{t1}}$. By hypothesis $R$ includes jet-correlated pairs $S$ (the jet signal) and (uncorrelated + $v_2$) background $B$. To isolate jet correlations $S$ background $B$ must be subtracted. The background form is assumed to be 
\bea \label{zyameq}
B(\Delta \phi;p_{t1},p_{t2})  &=& B_0+B_2(p_{t1},p_{t2})\, \cos(2\Delta \phi) \\ \nonumber
&=& B_0\, [1+2\, v_2(p_{t1})\, v_2(p_{t2})\, \cos(2\Delta \phi)],
\eea
where coefficient $B_0$ is determined by ZYAM and $v_2(p_{t1})\, v_2(p_{t2}) \rightarrow v_2^2$ is the actual value estimated from one or more $v_2^2\{{\rm method}\}$ measurements.

%%%%%%%%%%
 \begin{figure}[h]
  \includegraphics[width=3.3in,height=1.65in]{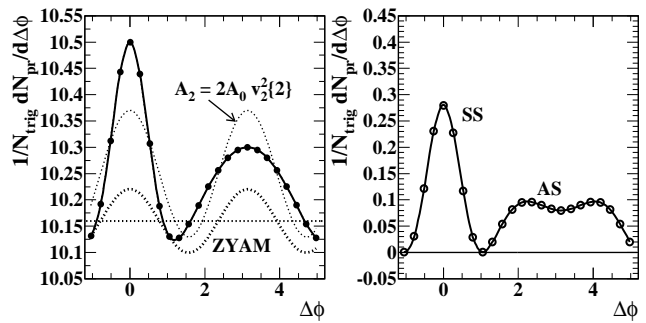}
\caption{\label{zyam}
Left: Simulated ``raw'' (unsubtracted) dihadron correlations (solid dots), azimuth quadrupole amplitude $A_2$ with corresponding sinusoid (light dotted curve) and ZYAM subtracted background (bold dotted curve). The ZYAM-estimated background offset is the dotted line.
Right: Result of ZYAM background subtraction in the left panel. The minimum at $\pi$ in the AS (away-side) peak is notable.
 } %nf28ab
 \end{figure}
%%%%%%%%%%%%

Fig.~\ref{zyam} (left panel) illustrates ZYAM subtraction. The points simulate a measured $R$ distribution (with quadrupole component $\equiv$ 0). The light dotted curve represents $v_2^2\{2\}$ obtained by fitting the entire distribution with $\cos(2\Delta \phi)$+constant (consistent with the definition of $v_2^2\{2\}$). The value of $v_2^2$ adopted for $B$ is $v_2^2\{2\}/2$, since $v_2^2\{4\} = 0$ holds for $R$ with no quadrupole component. The bold dotted curve is $B$ from Eq.~(\ref{zyameq}), with $B_0$ adjusted so that difference $S = R - B$ satisfies ZYAM. 

Fig.~\ref{zyam} (right panel) shows resulting ``jet'' correlations $S$ as the points. The ``raw'' jet component has been substantially reduced in amplitude, and the AS peak has a minimum at $\pi$. That is the general form of dihadron correlations for more-central A-A collisions after ZYAM subtraction. The solid curve is a free fit to the distribution discussed further below. The systematic error of the subtraction is sometimes estimated by the statistical error in a small interval about the minimum (e.g.~\cite{phenix}).

%%%%%%%%%
 \section{Azimuth correlation structure}

In p-p collisions triggered 1D azimuth correlations are approximated by Gaussian peaks centered at 0 and $\pi$ radians. 
In A-A collisions the structure evolves strongly with centrality. The AS peak broadens and a $\cos(2\Delta \phi)$ (quadrupole) sinusoid dominates the structure for intermediate centralities. Either jet peak may deviate from a Gaussian shape. The main question is how to isolate true jet correlation structure while minimizing bias.

\subsection{Periodic peak arrays}

Distributions on azimuth are periodic.  Peaks (nominally Gaussian) at 0 (SS, same-side) and $\pi$ (AS, away-side) are elements of periodic arrays. The SS array is centered on {\em even} multiples of $\pi$, the AS array on {\em odd} multiples. Nearby elements of an array outside a $2\pi$ interval can have a significant effect on distributions within the interval and should be included in fit models. Failure to include nearby members of both peak arrays may result in significant fitting errors, especially for broader peaks.

A peak array (SS or AS) can be represented by a  Fourier series of the form
\bea \label{fourier}
S(\Delta \phi;\sigma_{\Delta \phi},n) &=& A_{0,n} + A_{1,n}\,\{1 +  \cos(\Delta \phi - n\,\pi)\}/2 \nonumber  \\
&+& \sum_{m=2}^\infty A_{m,n}\, \cos(m\,[\Delta \phi - n\,\pi]),
\eea
where the $A_{m,n}$ are functions of Gaussian width $\sigma_{\Delta \phi}$, $n$ is even for SS peak arrays ($+$), and odd for AS arrays ($-$). The terms represent $2m$ poles, e.g. dipole ($m=1$), quadrupole ($m=2$), sextupole ($m=3$). As peak width $\sigma_{\Delta \phi}$ increases the number of significant terms in the series decreases. The limiting case is $\sigma_{\Delta \phi} \sim \pi / 2$, for which a peak array is approximated by a constant plus single dipole term---$[1+\cos(\Delta \phi)]/2$ (SS) or $[1 -\cos(\Delta \phi)]/2$ (AS). For Gaussian peak arrays with $\sigma_{\Delta \phi}< 1$ (and therefore nonzero sextupole) a Gaussian function is the more efficient representation.

\begin{figure}[h]
  \includegraphics[width=1.65in,height=1.65in]{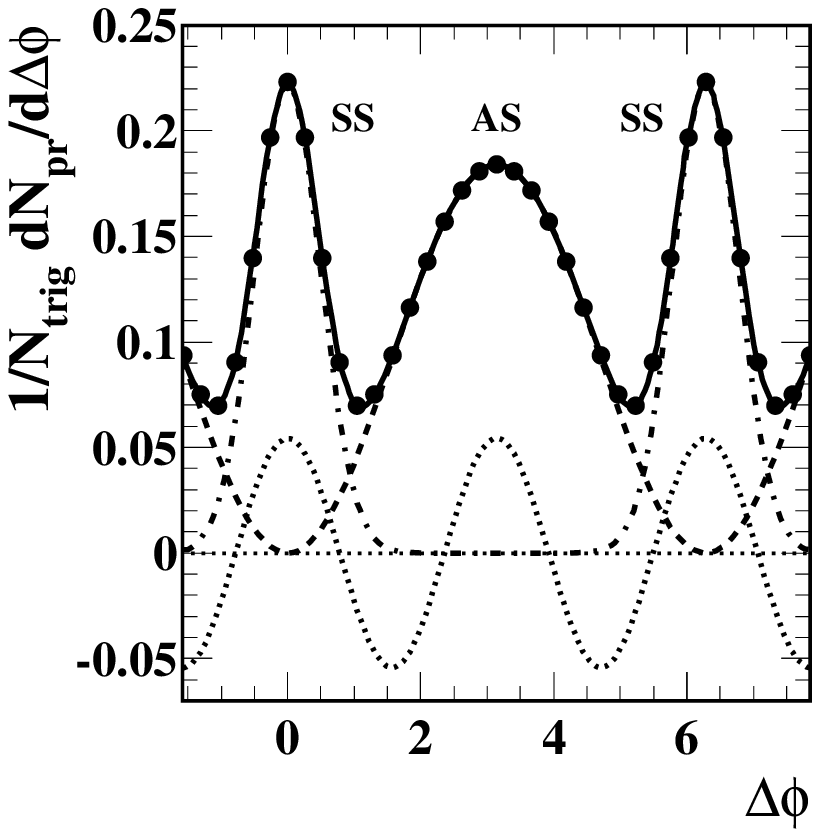}
  \includegraphics[width=1.65in,height=1.65in]{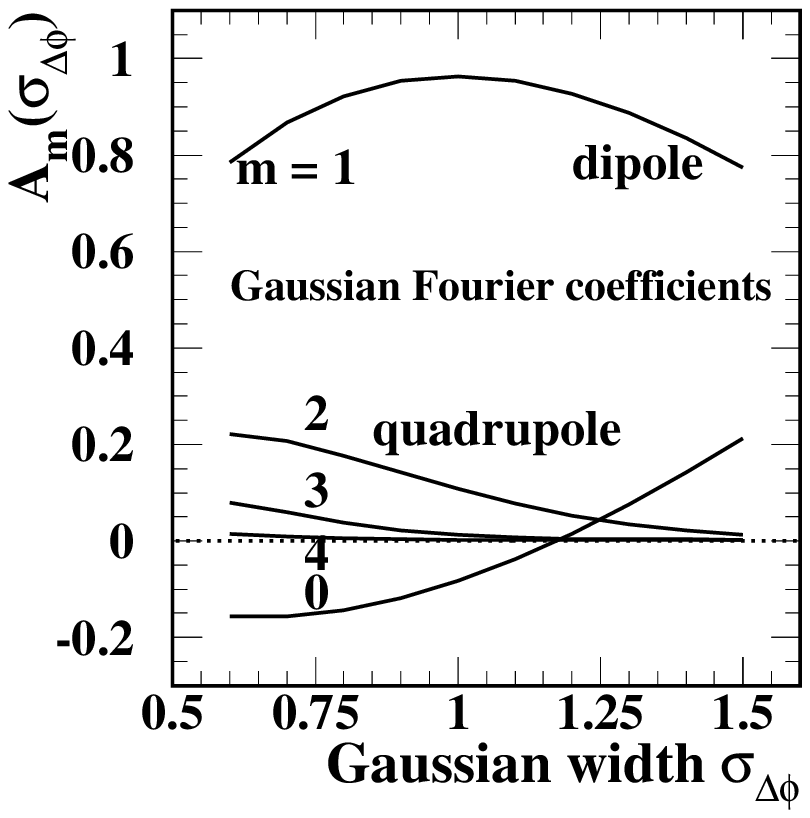}
\caption{\label{ortho}
Left: Periodic arrays of SS (same-side, dash-dotted) and AS (away-side, dashed) peaks. The SS peaks are Gaussians. The AS peaks are well-described by a dipole. The dotted sinusoid corresponds to the $m=2$ Fourier component of the SS peaks.
Right: Fourier amplitudes $A_m$ of a Gaussian [Eq.~(\ref{fourier})] vs peak width $\sigma_{\Delta \phi}$.
 } %nf30a, nf35a 
 \end{figure}

Fig.~\ref{ortho} (left panel) illustrates the sum of peak arrays (solid points) for SS and AS peaks extending beyond one $2\pi$ interval. The SS Gaussian peak array is the dash-dotted curve, the AS array with $\sigma_{\Delta \phi} \sim \pi / 2$ is the dashed curve (approximately pure dipole). The dotted curve is the quadrupole term of the SS array, which would add a large ``nonflow'' contribution $\delta_{2,+}$ to $v^2_2\{2\}$ inferred from that distribution (see next subsection). 

%%%%%%%%%%
%%%%%%%%%%%%

Fig.~\ref{ortho} (right panel) shows the Fourier amplitudes of a Gaussian peak for the first five terms of Eq.~(\ref{fourier}) as functions of the r.m.s.~peak width. For $\sigma_{\Delta \phi} \sim \pi/2$ only the constant and dipole terms contribute. For narrower peaks terms with $m > 1$ become significant. 

\subsection{The quadrupole component}

Analysis whose main objective is ``elliptic flow'' measures some part of the total quadrupole component of azimuth correlations depending on the specific ``method.'' There are three phenomenological sources of the total quadrupole component: the SS jet peak, the AS jet peak and the non-jet (NJ) background. The NJ component may or may not be related to a hydro phenomenon. 

The $m=2$ Fourier amplitude of the entire azimuth distribution is $A_2 = 2\, A_0\,v^2_2\{2\}$, based on the definition of $v^2_2\{2\} \sim v^2_2\{EP\}$. The $m=2$ Fourier terms from the SS and AS peak arrays contribute to $v^2_2\{2\}$ a combined ``nonflow'' systematic error $\delta_2$. If the AS peak is broad (typical for more-central A-A collisions and/or untriggered correlations) ``nonflow'' $\delta_2$ is dominated by the SS peak.
The following correspondence can be made to conventional terminology and symbols. $A_{2,\pm} \equiv 2\, A_{0,\pm}\, \delta_{2,\pm}$. $\delta_{2,+} + \delta_{2,-} = \delta_2$ (conventional ``nonflow'' measure). $v_2^2\{EP\} \approx v_2^2\{2\} = \delta_2 + v_2^2\{NJ\}$. $v_2^2\{2D\} = \delta_{2,-} + v_2^2\{NJ\}$ (from untriggered 2D autocorrelations). For typical untriggered correlations $\delta_{2,-} \ll \delta_{2,+}, v_2^2\{NJ\}$ and $v_2^2\{2D\} \approx v_2^2\{NJ\}$. Background component $B_2 = 2\, B_0\, v_2^2\{NJ\}$ may or may not be caused by a hydro mechanism~\cite{gluequad}.

\subsection{Peak models} \label{pythia}

Concern has been expressed that the background offset cannot be determined accurately by peak model fits if the exact form of the jet peaks is not known {\em a priori}. It is claimed that due to in-medium modifications peak shapes may not be Gaussian. ``Rigorous decomposition of the jet from its underlying event currently requires assumptions about the jet shape or the physics of the underlying event''~\cite{phenix}. Studies with PYTHIA are interpreted to conclude a substantial difference between peak shapes with and without radiation effects, including non-Gaussian shapes. Given such difficulties the ZYAM procedure is invoked as an alternative.

%%%%%%%%%%
 \begin{figure}[h]
  \includegraphics[width=2.8in,height=1.6in]{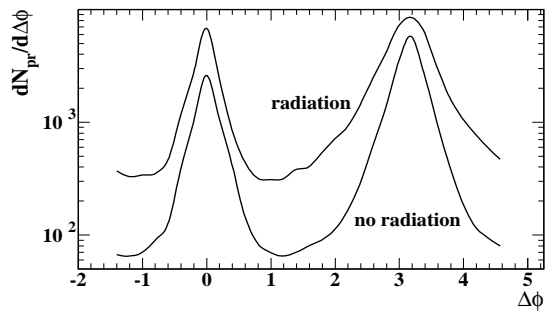}
\caption{\label{pyth}
Simulated dihadron correlations for p-p collisions from PYTHIA~\cite{phenix} with cuts 3-5$\times$3-5 GeV/c applied illustrating deviations from Gaussian peak structures with (upper) and without (lower) initial- and final-state radiation invoked.
 } %nf38e \\
 \end{figure}
%%%%%%%%%%%%

Fig.~\ref{pyth} shows PYTHIA simulations (samples from Fig.~5 of~\cite{phenix}) which reveal changes in peak shapes with and without initial- and final-state radiation. The simulation is invoked to support the conclusion that jet peak shapes are sufficiently uncertain (i.e., non-Gaussian) that peak models cannot be used to determine the background offset, that the alternative ZYAM method is {\em necessary}.

To test that assertion the modified-Gaussian form
\bea \label{blah}
\hat G(\Delta \phi;\sigma_{\Delta \phi},n) &=& \frac{A}{\left\{1 + \frac{1}{2n}\left(\frac{\Delta \phi}{\sigma_{\Delta \phi}}\right)^2\right\}^n}
\eea
 is compared to data. That expression is to Gaussians what the L\'evy distribution is to exponentials~\cite{levy}. The L\'evy distribution, the average of an ensemble of exponential distributions on variable $x$ with fluctuating slope parameter $X$ (e.g., $m_t,T$ for Maxwell-Boltzmann distributions), has the form $1/(1+x/X\, n)^n$, where $1/n = \sigma^2_X / \bar X^2$ represents the relative variance of $X$ over the ensemble. Eq.~(\ref{blah}) describes an ensemble of Gaussians with fluctuating widths $\sigma_{\Delta \phi}$ and relative variance $ \sigma^2_{\sigma_{\Delta \phi}} /  \sigma_{\Delta \phi}^2 = 1/4n$. Limiting cases for $1/n \rightarrow 0$ are $\exp(-x/X)$ and $\exp\{-(\Delta \phi/\sigma_{\Delta \phi})^2/2\}$.

\begin{figure}[h]
  \includegraphics[width=3.3in]{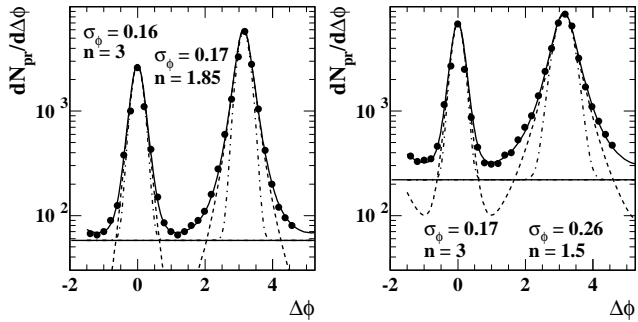}
\caption{\label{gauss1}
Left: Data with no radiation from Fig.~\ref{pyth} (points) described by modified Gaussians as in Eq.~(\ref{blah}) plus a constant background (solid curve). The peak parameters are noted. Dashed curves are with no background. Dash-dotted curves are with no width fluctuations ($1/4n = 0$).
Right: Same, but for data with radiation invoked in PYTHIA.
 } %nf38ab
 \end{figure}

Figure~\ref{gauss1} shows the samples in Fig.~\ref{pyth} as points. The solid curves through points are $\hat G_{SS} + \hat G_{AS}$ + constant. The $\hat G$ parameters are noted and background constants are indicated by the solid lines. The solid curves accurately represent the simulations. The background offsets are well-determined. The dashed curves omit the backgrounds. The dash-dotted curves include $\hat G_{SS}$ and $\hat G_{AS}$ with $1/n \rightarrow 0$---pure Gaussians $G$ with the same widths. 

%%%%%%%%%%
%%%%%%%%%%%%

\begin{figure}[h]
  \includegraphics[width=3.3in]{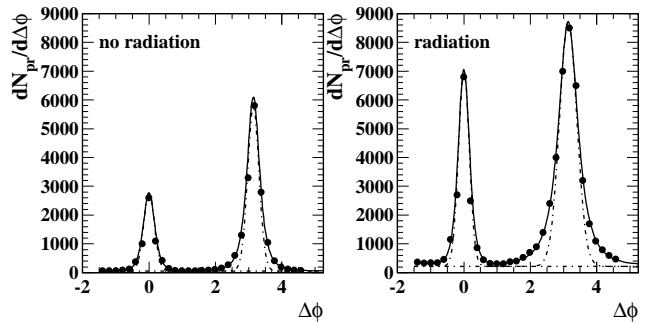}
\caption{\label{gauss2}
Same as Fig.~\ref{gauss1} but plotted on a linear scale.
 } %nf38cd 
 \end{figure}

Figure~\ref{gauss2} shows the same elements on a linear format. Especially for the SS peaks the difference between true and modified Gaussian is subtle.

%%%%%%%%%%
%%%%%%%%%%%%

The good agreement between Eq.~(\ref{blah}) and the PYTHIA simulations reveals that deviations from Gaussians are primarily due to width fluctuations. For the extreme case of Fig.~\ref{pyth} (high $p_t$ cuts, $\sigma_{\Delta \phi} \sim 0.15$) deviations of the peak shape from a Gaussian in p-p collisions are obvious on a semilog plot. However, the peak shape is amenable to exact description with a simple extension of the Gaussian model function to include fluctuations. 

For cut systems typically encountered in dihadron studies at RHIC the SS peak width is more typically 0.4 to 0.5 rather than 0.16 as in the PYTHIA study. Other things being equal that implies a {\em ten-fold decrease} in relative variance $1/4n$. SS peak shapes should then be indistinguishable from true Gaussians. In central Au-Au collisions fluctuations in the SS peak width should be reduced five-fold by central-limit averaging, since there are on average $30\pm 6$ minijets in each collision~\cite{fragevo}. The mean AS peak width increases substantially in more-central Au-Au collisions, further reducing the effective $1/4n$, but driving the peak shape toward a pure dipole shape.

For p-p collisions and $p_t$ cuts below 6 GeV/c, especially for minijet (minimum-bias jet) structure, and for more-central Au-Au collisions, the SS peak shape is accurately modeled by a Gaussian. The AS peak is accurately modeled by three terms of a Fourier series (or a Gaussian in some cases). Background subtraction with small systematic uncertainties can be achieved by peak fitting.

 \subsection{2D angular autocorrelations} \label{2dauto}

2D angular autocorrelations on difference variables $\eta_\Delta = \eta_1 - \eta_2$ and $\phi_\Delta = \phi_1 - \phi_2$ provide substantial additional information beyond the 1D azimuth projection. The typical correlation structure includes a 2D SS peak, an AS peak on azimuth (AS ridge) independent of pseudorapidity difference, and an azimuth quadrupole~\cite{newflow,gluequad}. 

The SS 2D peak is unambiguously isolated by model fits because of its variation on pseudorapidity difference. The AS ridge and azimuth quadrupole are uniform on $\eta_\Delta$ and thus project onto 1D azimuth without loss of information. The AS peak is typically dominated by the dipole term. For more-central A-A collisions, and generally for untriggered correlations, the AS peak may be pure dipole, in which case it is orthogonal to the quadrupole component. For peripheral A-A or p-p collisions and higher-$p_t$ cuts an AS Gaussian is more efficient. For intermediate cases ($\sigma_{\phi_\Delta} \sim 1.0$ - 1.2) the AS dipole model may allow a small contribution $\delta_{2,-}$ to measured quadrupole amplitude $v_2^2\{2D\}$ which then establishes an {\em upper limit} on any independent (nonjet) quadrupole component $v_2^2\{NJ\}$.

\begin{figure}[h]
  \includegraphics[width=1.65in,height=1.5in]{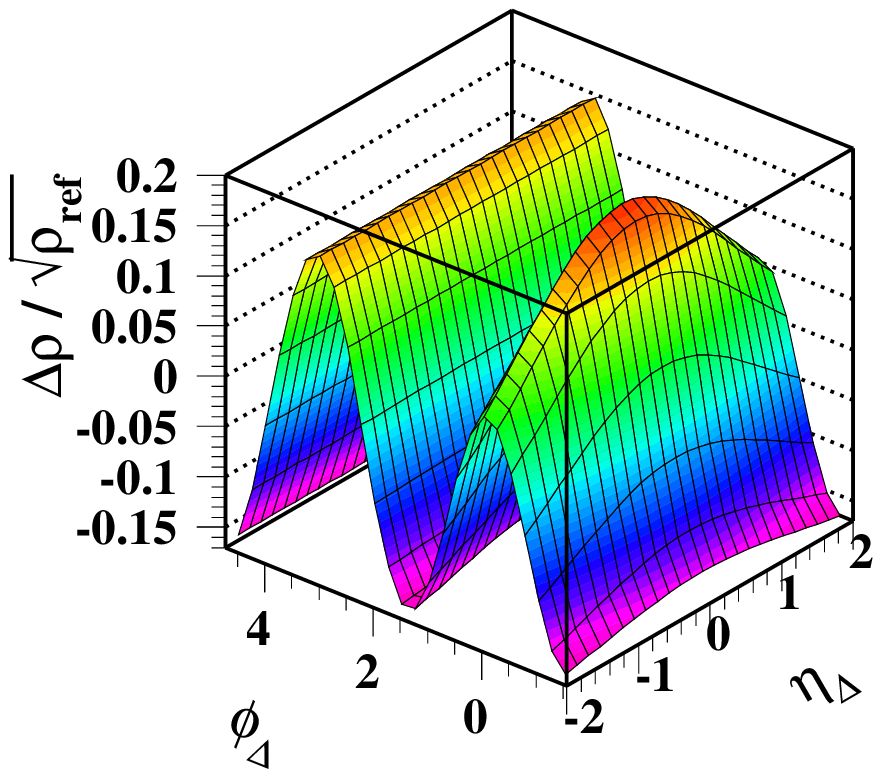}
  \includegraphics[width=1.65in,height=1.5in]{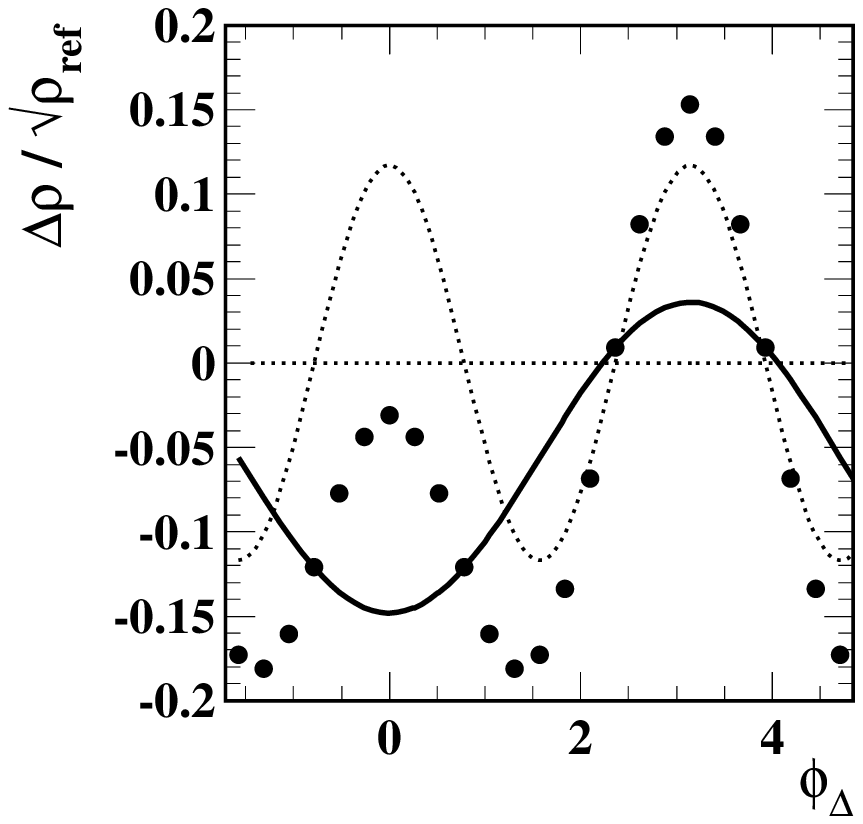}
\caption{\label{ortho2}
Left: Simulated 2D angular autocorrelation showing an $\eta$-elongated SS peak, an AS ridge and quadrupole.
Right: The 2D autocorrelation with SS peak subtracted and projected onto $\phi_\Delta$ (points). The result is a dipole (solid curve) plus quadrupole (dotted curve).
 } %nf39, nf30b 
 \end{figure}

Fig.~\ref{ortho2} (left panel) illustrates untriggered (no special $p_t$ cuts)  2D angular correlations for 200 GeV mid-central Au-Au collisions~\cite{daugherity}. The SS peak is a 2D Gaussian which can be reliably isolated from the other correlation structure. The remaining structure then consists of $[1-\cos(\phi_\Delta)]/2$ dipole and $\cos(2\phi_\Delta)$ quadrupole components.

%%%%%%%%%%
%%%%%%%%%%%%

Fig.~\ref{ortho2} (right panel) shows the distribution in the left panel with the 2D Gaussian fit removed and the difference projected onto azimuth (points). The AS peak width for untriggered correlations is typically $\sigma_{\phi_\Delta} \sim \pi / 2$, and the peak is essentially a dipole. The AS peak and quadrupole are therefore orthogonal, implying negligible covariances among the three model components: SS 2D Gaussian, AS dipole and quadrupole.

2D autocorrelations can also be constructed for specific (symmetric or asymmetric) $p_t$ cuts, providing combinatorial triggered jet reconstruction with reduced ambiguity. 2D  histograms provide the most direct and accurate access to joint distribution $v_2^2(p_{t1},p_{t2})$. There is no evidence in 2D autocorrelations for the AS double-peaked structures resulting from ZYAM subtraction.

%%%%%%%%%
 \section{ZYAM Critique}

The purpose of this analysis is to examine the validity of the ZYAM procedure and consequent novel structures in reported jet correlations. In this section we review the logic and consequences of ZYAM. 

\subsection{Assumptions supporting ZYAM}

The background level (zero offset) of sibling-pair distribution $R$ in Eq.~(\ref{raw}) is well-defined when SS and AS peaks are completely separated, as in p-p collisions with ``high-$p_t$'' cuts. For other collision systems or lower $p_t$ cuts the peaks may overlap, and the background level is not then visually obvious. For overlapping peaks and at higher $p_t$, especially in more-central A-A collisions, $v_2^2$ subtraction is also ambiguous.

The ZYAM procedure explicitly assumes that  subtracted background $B$ emerges from a thermalized bulk medium. Almost all particles, otherwise uncorrelated, are then correlated with the reaction plane and described by $v_2^2$, motivating the structure of $B$ in Eq.~(\ref{zyameq}). The problem remains to determine the true background offset for azimuth correlations, the constant $B_0$ in Eq.~(\ref{zyameq}). In principle, the background for sibling pairs accepted in cut box $(p_{t1},p_{t2})$ should be the mean pair ratio $\langle n_1\, n_2 \rangle / \bar n_1\, \bar n_2$ times the mixed-pair background normalized to pair number $\bar n_1\, \bar n_2 $, in which case $B_0 = \langle n_1\, n_2 \rangle$. 

However, covariance $\langle n_1\, n_2 \rangle - \bar n_1\, \bar n_2$ (a fluctuation measure) contains contributions other than jet correlations. Positive contributions include centrality ($b$) fluctuations and correlations with characteristic lengths exceeding the angular acceptance. Negative contributions include possible local anticorrelations and, most importantly, canonical suppression of fluctuations if a centrality condition is imposed on $n_{ch}$ within the pair angular acceptance or some part of it.  Thus, $B_0$ cannot be determined directly from present fluctuation [$\langle n_1\, n_2 \rangle$] measurements.

Alternatively, $B_0$ could be determined directly from correlation structure by model (e.g., Gaussian) fits to the SS and AS peaks. The main argument against model fitting is that the collision process could modify peak structures away from a  Gaussian shape. Incorrect fit models might then lead to biased background subtraction. If SS and AS peaks are overlapping ZYAM is invoked by that argument to determine $B_0$.

ZYAM implicitly assumes that real distributions do not contain overlapping peaks, which contradicts many examples in nature including nuclear collisions. For most of the centrality range of Au-Au collisions and some $p_t$ cuts the same-side and away-side peaks do overlap strongly, and the true background offset can differ greatly from the ZYAM estimate. 

 \subsection{Estimating $\bf v_2^2(p_{t1},p_{t2})$}

The subtracted background includes $v_2^2$ correlations (azimuth quadrupole). In Eq.~(\ref{zyameq}) joint distribution $v_2^2(p_{t1},p_{t2})$ has been replaced by product $v_2(p_{t1})\, v_2(p_{t2})$ assuming factorization. The factors (marginal distributions) are determined by separate analysis. Factorization is defended by the claim that in a thermalized bulk medium particles have no correlations among themselves, only with the common reaction plane.

$v_2(p_t)$ is measured by several methods denoted by $v_2$\{\text{method}\}, e.g., $v_2\{2\}$ (two-particle correlations), $v_2\{EP\}$ (event plane), $v_2\{4\}$ (four-particle cumulants), etc.~\cite{flowmeth}. At larger $p_t$ $v_2\{2\}$ or $v_2\{EP\}$ is favored because of smaller particle yields. Both are strongly sensitive to minijets (nonflow). Extraction of an azimuth quadrupole component from 2D angular correlations provides alternative measure $v_2\{2D\}$ which is by construction independent of jet correlations to good approximation~\cite{gluequad}.

In conventional flow analysis it has been assumed that $v_2^2\{2\} = \bar v_2^2 + \sigma^2_{v_2}$ and  $v_2^2\{4\} = \bar v_2^2 - \sigma^2_{v_2}$, so that mean value $\left(v_2^2\{2\} + v_2^2\{4\} \right)/2 = \bar v_2^2$ eliminates ``flow fluctuations'' $\sigma^2_{v_2}$~\cite{flowflucts, gluequad}. The mean value is often used to determine $v_2(p_t)$ factors for Eq.~(\ref{zyameq}). But $v_2^2\{2\} = v_2^2\{2D\} + \delta_2$ which defines ``nonflow'' $\delta_2$, dominated by the $m=2$ Fourier component of the SS Gaussian (mini)jet peak~\cite{gluequad}. And $v_2^2\{4\}$ may also include minijet contribution $\delta_2'$ in more-central Au-Au collisions, because minijets then have multiplicities $\geq 4$. Thus, $\left(v_2^2\{2\} + v_2^2\{4\} \right)/2 \rightarrow v_2^2\{2D\} - \sigma^2_{v_2} +  (\delta_2 +  \delta_2') / 2$, and independent (nonjet) quadrupole component $v_2^2\{2D\}$ may be a small fraction of the nonflow (jet) terms for more-central Au-Au collisions.

In Fig.~\ref{zyam} (left panel) the light dotted curve represents $v_2^2\{2\}$ obtained (per its definition) by fitting the entire azimuth distribution with $\cos(2\Delta \phi)$. By construction the data contain no independent quadrupole component. Thus,  $v_2^2\{2\} = \delta_2$ is entirely ``nonflow'' or jet correlations. In essence, the quadrupole component of jet correlations is subtracted from jet correlations to produce distortions. 

\subsection{A typical ZYAM analysis}

\begin{figure}[h]
  \includegraphics[width=3.3in,height=1.65in]{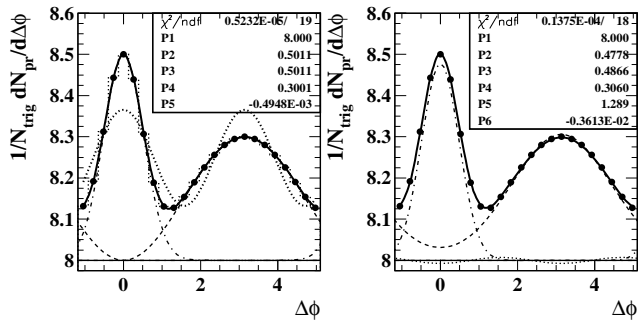}
\caption{\label{starnewa}
Left: Simulated azimuth correlation data (points) including a SS Gaussian (dash-dotted curve) and AS dipole (dashed curve). The bold solid curve is a free fit to the data which returns the model parameters: offset (P1), SS Gaussian amplitude and width (P2, P3), dipole amplitude (P4), quadrupole amplitude (P5).
Right: The same data fitted with offset and SS Gaussian (as before), AS Gaussian amplitude and width (P4, P5) and quadrupole (P6).
 } %nf33ab \\
 \end{figure}

Figure~\ref{starnewa} (left panel) shows simulated azimuth correlation data (points) for central ($b=0$) Au-Au collisions. The distribution consists of SS Gaussian and AS dipole only. A fit to unsubtracted data using a Gaussian+dipole+quadrupole+constant model (solid curve) by construction returns the simulation parameters. Also shown is a fit to the distribution with $A_0 +A_2\,\cos(2\Delta \phi)$ only (dotted curve), which for this case corresponds to $A_2 \equiv 2A_0\, v^2_2\{2\} = 0.12 \simeq  P_2/4$ by definition of $v^2_2\{2\}$.

%%%%%%%%%%
%%%%%%%%%%%%

Figure~\ref{starnewa} (right panel) shows a fit to the same ``data'' with two Gaussians and a quadrupole. In comparing the two panels several aspects are notable. 1) The offset levels are the same. 2) The properties of the SS Gaussian agree to a few percent. 3) The amplitude of the AS peak agrees to a percent. 4) The inferred quadrupole amplitude (zero in the simulation) is, in either case, 1\% or less of the dipole amplitude. The inferred r.m.s.~width 1.3 of the AS Gaussian is consistent with the dipole model.

\begin{figure}[h]
  \includegraphics[width=3.3in,height=1.63in]{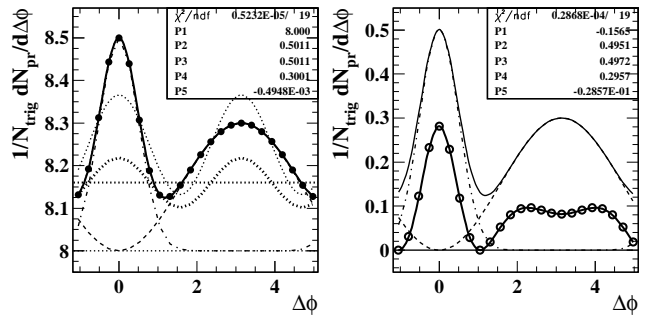}
\caption{\label{starnewb}
Left: Simulated data as in Fig.~\ref{starnewa} with ZYAM background model (bold dotted curve).
Right: Result of subtracting ZYAM background model from simulated data (open symbols). The bold solid curve is a free fit to the subtracted data which returns the original SS Gaussian and AS dipole model parameters plus the ZYAM offset and quadrupole parameters.
 } %nf26ab
 \end{figure}

Figure~\ref{starnewb} (left panel) uses the same simulated data to illustrate the ZYAM procedure. For this example $v_2^2\{2\}$ is determined as above, and (ideal) $v^2_2\{4\}$ is zero. If the $v_2^2$ subtraction is defined by quadratic mean $v_2^2 = (v_2^2\{2\}+v^2_2\{4\})/2$ then since $A_0 \sim 8$ the subtracted value is $v_2^2 =0.12/(2\times 2\times 8)$, or $v_2 = 0.06$ (compared to 0.07 in~\cite{staras}). 

%%%%%%%%%%
%%%%%%%%%%%%

Figure~\ref{starnewb} (right panel) shows the result of ZYAM subtraction as the points. The dash-dotted and dashed curves show the simulation input, the ``right answer.'' The bold solid curve through the points is a free fit with the SS Gaussian+AS dipole+quadrupole model. The original parameters are returned in addition to the ZYAM-imposed $v_2^2$ and offset. In other words, no distribution information is lost by the ZYAM subtraction, but the result is visually misleading. In particular, the $v_2^2$ oversubtraction results in a minimum in the AS peak, and the ZYAM offset subtraction imposes a large reduction in {apparent} jet yields. The same fit model was applied in Fig.~\ref{zyam} (right panel), with similar results.

\subsection{Consequences of ZYAM}

ZYAM subtraction with overestimated $v_2$ values results in distorted azimuth correlations, including a misleading double-peaked away-side structure conventionally interpreted in terms of ``Mach cones''~\cite{machth1,machth2}. Oversubtraction of $v_2^2$ occurs for three reasons: 1) $v_2^2\{2\}$ (or $v_2^2\{EP\}$) includes large ``nonflow'' (minijet) contributions for all A-A centralities but especially for more-central collisions, 2) $v_2^2\{4\}$ can also include minijet contributions for more-central collisions and 3) factorization of the joint $v_2^2(p_{t1},p_{t2})$ distribution can be questioned given asymmetric (off-diagonal) $p_t$ cuts.

Fig.~\ref{starnewb} (right panel) shows nominal jet signal $S = R - B$ as the points, illustrating the general form of more-central dihadron correlations after ZYAM subtraction. The original jet correlations are strongly reduced, and the away-side peak has a minimum at $\pi$. The apparent reduction of jet correlations appears to confirm strong jet quenching and parton thermalization. The AS peak structure also suggests parton interaction with a dense, thermalized medium leading to shock-wave formation.

The form of the ZYAM-subtracted ``data'' in Fig.~\ref{starnewb} (right panel) depends only weakly on collision conditions, because $v_2^2\{2\}$ is dominated by the $m=2$ Fourier component of the SS peak. Its magnitude is therefore locked to the jet-peak amplitude, and all correlation structure scales up and down together with collision centrality. 

\begin{figure}[h]
   \includegraphics[width=1.65in,height=1.65in]{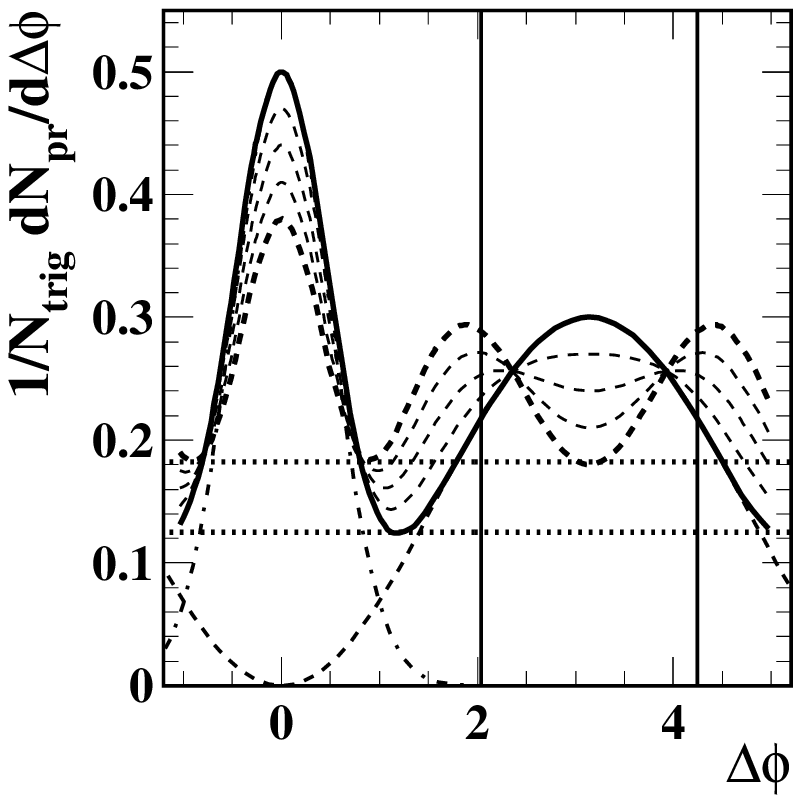}
 \includegraphics[width=1.65in,height=1.65in]{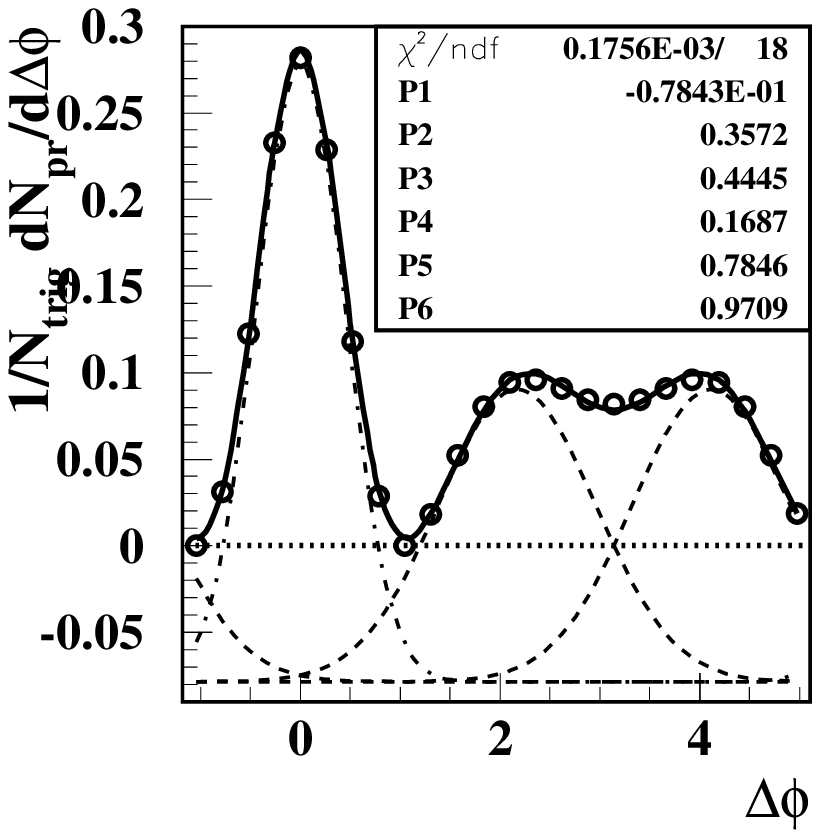}
\caption{\label{peakfit}
Left: Demonstration of the effects of $v_2$ oversubtraction. Model jet correlations from Fig.~\ref{starnewa} are the ``data'' (solid curve) from which quadrupole components of successively larger amplitudes are subtracted (dashed curves). Dotted lines represent zero levels established according to the ZYAM principle. Vertical solid reference lines are at $\pi \pm 1.1$.
Right: ZYAM-subtracted distribution from Fig.~\ref{starnewb} (right panel) fitted with single SS Gaussian and two AS Gaussians located symmetrically about $\pi$. Common AS Gaussian amplitudes and widths are P4 and P5. Displacements from $\pi$ are P6.
 }  %nf40b, nf40a \\
 \end{figure}

Figure~\ref{peakfit} (left panel) demonstrates that maxima near $\Delta \phi \approx \pi \pm 1$ resulting from ZYAM subtraction are an inevitable consequence of $v_2$ oversubtraction for typical angular correlations from RHIC collisions. For a generic SS/AS peak combination in more-central A-A collisions several values of $A_2$ (integer multiples of $2A_0\, v_2^2 = 0.015$) including zero (solid curve) are invoked in the subtraction. The persistence of apparently-displaced peaks at the same locations is evident. Note the effect of ZYAM subtraction even for $v_2^2 = 0$ (solid curve and dotted line).

%%%%%%%%%%
%%%%%%%%%%%%

Figure~\ref{peakfit} (right panel) illustrates attempts to characterize the AS peak structure with two Gaussians~\cite{phenix}. Nominal medium-induced ``shoulder'' Gaussians are located near $\Delta \phi \approx \pi \pm 1$. In some cases a fragmentation-related ``head'' Gaussian is introduced at $\Delta \phi  = \pi$. The bold solid curve is a free fit with shoulder Gaussians allowed to vary in centroid, width and amplitude, but symmetrically about $\Delta \phi = \pi$. The detailed model fit can be contrasted with the light (not bold) solid curve of Fig.~\ref{starnewb} (right panel), the ``right answer'' for these ``data.'' 

In Figure~\ref{peakfit} (right panel) the offset is allowed to vary, and the SS Gaussian amplitude is 0.36, closer to input value 0.50. If the offset were fixed at zero the SS peak amplitude would be 0.29, as in Fig.~\ref{starnewb} (right panel). If the correct fit model were used the original simulation parameters would be returned. This exercise reveals the typical magnitude of systematic errors resulting from ZYAM and $v_2^2$ oversubtraction.

\subsection{Systematic errors and uncertainties}

ZYAM subtraction uncertainties derive from uncertainties in offset $B_0$ and $v_2^2$. The ZYAM offset uncertainty is assumed to result only from determining the effective minimum of the unsubtracted (raw) pair distribution, which is in turn assumed to be dominated by the statistical error within a small azimuth interval~\cite{starprl,phenix}.

The present analysis demonstrates that the ZYAM assumption is contradicted by {any} distribution with overlapping peaks. The true ZYAM systematic uncertainty is indicated by the typical difference between a ZYAM offset value and that inferred from peak fitting as in this analysis. The error can be large (up to 100\% of true jet correlation amplitudes), leading to substantial underestimation of jet correlations and related hadron yields.

$v_2^2$ subtraction uncertainties are dominated by measurement uncertainties for marginal $v_2^2(p_t)$ distributions. Especially for more-peripheral or -central A-A collisions minijet contributions to $v_2^2$ typically persist as large uncorrected systematic errors. $v_2^2(p_t)$ is estimated by average $(v_2^2\{2\}+v_2^2\{4\})/2$~\cite{flowflucts}. Where $v_2^2\{4\}$ data are not available (e.g., 0-5\% central Au-Au) the approximation $v^2_2\{4\} \sim v^2_2\{2\}/4$ is sometimes made~\cite{starprl}. The {\em minimum} value of $v_2^2$ (e.g., for central Au-Au collisions) is then $v_2^2\{2\}/2$ if (ideal) $v_2\{4\}$ is known (i.e., zero) or $5\, v_2^2\{2\}/8$ if $v_2\{4\}$ is not known. 

In central ($b = 0$) A-A collisions $\bar v_2^2 = 0$ and $v_2^2\{2\} = \delta_2$. Thus, the minimum value of $v_2^2$ assumed for background subtraction in central collisions is 1/2 the $m=2$ Fourier component of the SS jet peak. Jet correlations are converted to ``nonflow'' which is subtracted from jet correlations. Consequently, jet yields can be greatly underestimated.

%%%%%%%%%
 \section{ZYAM examples from RHIC data}

Two examples from ZYAM-based jet correlation analysis of RHIC 200 GeV Au-Au data are compared. ZYAM background subtraction is reversed to recover nominally-undistorted jet correlations. Free fits to data comparing a Gaussian-dipole-quadrupole model with a Gaussian-Gaussian-quadrupole model reveal consistent jet correlations in the two cases.

\subsection{Dihadron correlations from STAR}

\begin{figure}[h]
  \includegraphics[width=1.65in]{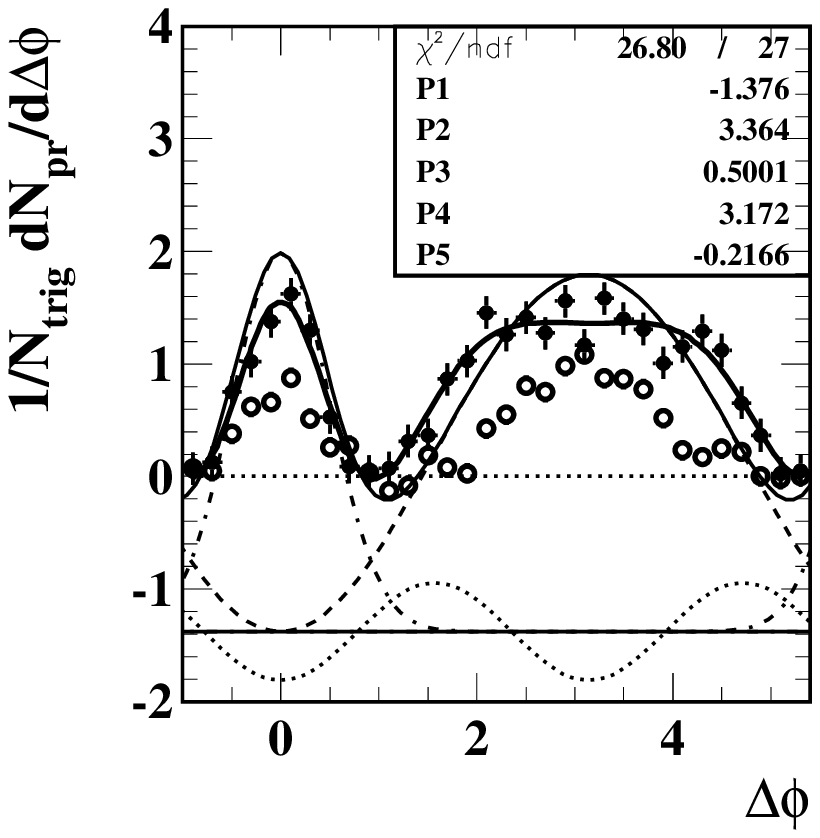}
  \includegraphics[width=1.65in]{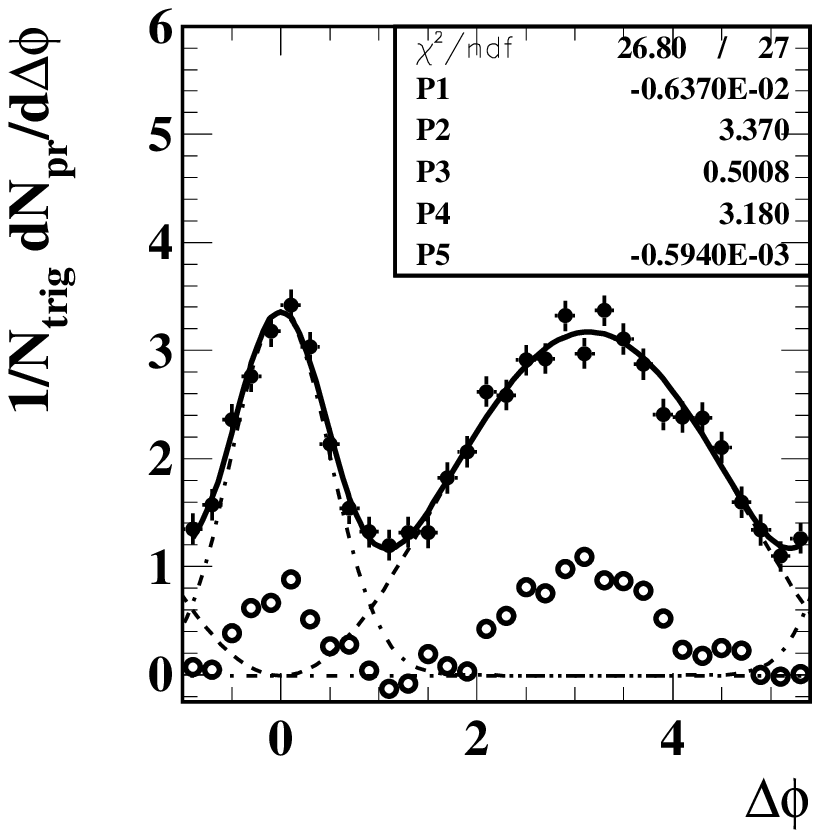}
  \includegraphics[width=1.65in]{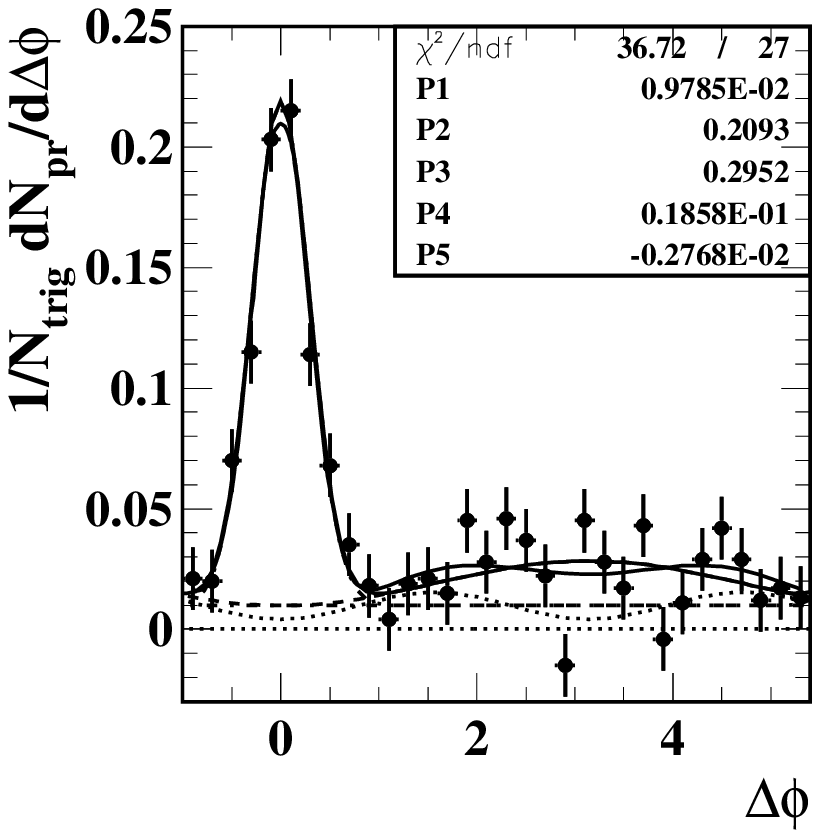}
  \includegraphics[width=1.65in]{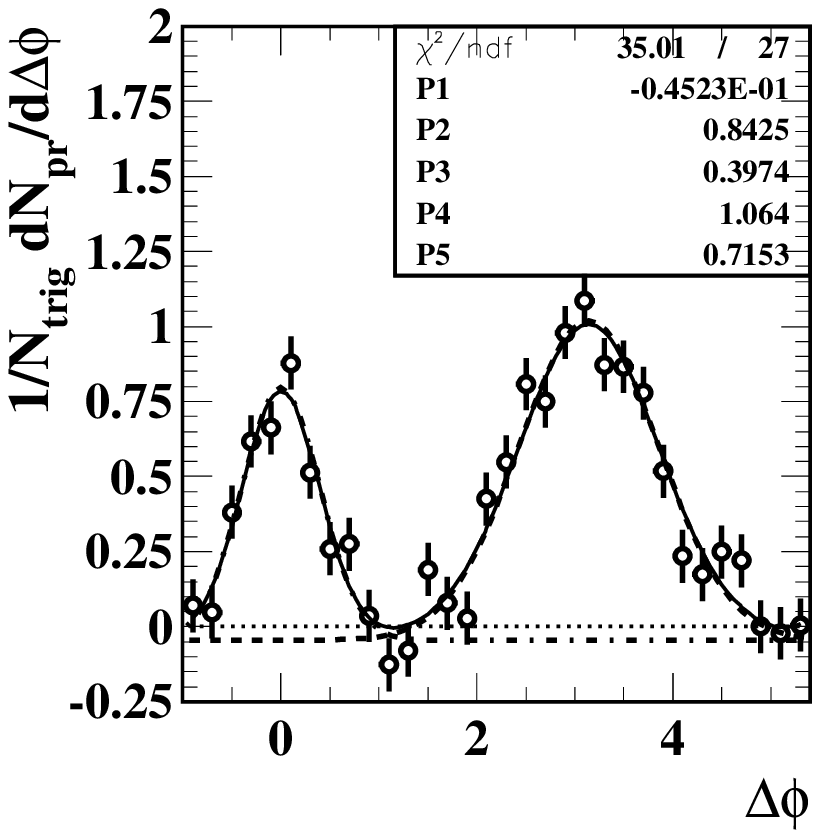}
\caption{\label{starold}
Upper left: ZYAM-subtracted angular correlations for 0-12\% central 200 GeV Au-Au collisions (solid points)~\cite{starprl} with free fit (bold solid curve) of SS Gaussian (dash-dotted curve), AS dipole (dashed curve) and quadrupole (dotted curve). Open points are p-p data relative to ZYAM zero.
Upper right: ZYAM subtraction reversed, true zero level recovered from free fits to data (solid points), compared to p-p data (open symbols).
Lower left: Consequences of raising the associated $p_t$ cut to 2 GeV/c~\cite{starprl}. Most of the SS fragment distribution is rejected and AS correlations are strongly suppressed.
Lower right: Fit of two Gaussians plus offset to p-p data.
 } %nf34 \\
%Add (a), (b) etc
 \end{figure}

Fig.~\ref{starold} (upper left) shows data from Fig.~1 (upper left) of~\cite{starprl}. ZYAM-subtracted data from 200 GeV 0-12\% central Au-Au collisions for (trigger$\times$associated) 4-6$\times$0.15-4 GeV/c $p_t$ cuts are shown as solid points. Corresponding p-p data are shown as open circles. The bold solid curve is a free fit with offset (P1), SS Gaussian (amplitude P2, width P3), AS dipole (P4) and quadrupole (P5). The resulting offset is the solid line at $P_1$, the SS Gaussian is the dash-dotted curve, the dipole is the dashed curve, and the (negative) quadrupole is the dotted curve. 

Fig.~\ref{starold} (upper right) shows a reconstruction of the original (``raw'') data distribution prior to ZYAM subtraction based on $P_1$ and $P_5$. The relation of p-p to central Au-Au data is quite different. Both SS and AS peaks {\em increase by a factor six} from p-p to central Au-Au, a large increase in the jet yield which is not apparent in the upper-left panel determined by conventional ZYAM subtraction.

%%%%%%%%%%
%%%%%%%%%%%%

 Without the unsubtracted data we cannot determine the original quadrupole component exactly. However, from information in the upper-right panel we can estimate what value of $2\, A_0\, v_2^2$ was used in the subtraction. The SS peak amplitude is $P_2 = 3.37$ and the width is $P_3 =0.50$. Fig.~\ref{ortho} (right panel) indicates that $m=2$ amplitude $A_2$ for that width is 25\% of the SS amplitude. Using the prescription $v_2^2 = (v_2^2\{2\} + v_2^2\{4\})/2 \rightarrow v_2^2\{2\}/2$ for central collisions the minimum value of $2\,A_0\, v_2^2 = 0.25\times 3.37/2 = 0.42$ from the definition of $v_2^2\{2\}$. The corresponding number from ZYAM subtraction in the upper-left panel is $-2\,P_5 = 0.42$, suggesting that the quadrupole component of the unsubtracted data was consistent with zero. 

Fig.~\ref{starold} (lower left) shows data from Fig.~1 (lower left) of~\cite{starprl} with the same fit model applied but with the associated $p_t$ cut raised to 2-4 GeV/c. While the AS peak is strongly reduced by the elevated associated-$p_t$ cut it is still described by a dipole to the error limits of the data.

Fig.~\ref{starold} (lower right) shows a two-Gaussian fit to the p-p data from~\cite{starprl}. Comparing p-p and Au-Au data the SS peak width changes only slightly from p-p (0.4) to central Au-Au (0.5). The AS peak width doubles from p-p (0.7) to Au-Au ($\sim 1.4$). The latter should be expected due to increased $k_t$ broadening of the AS peak, since the mean participant-nucleon path length (as N-N binary collisions) increases from 1 to 6 in central Au-Au~\cite{centmeth}.

Figure 2 (upper panel) of~\cite{starprl} reviews the apparent centrality systematics of SS and AS peak integrals. In either case an increasing trend for peripheral collisions appears to saturate for more-central collisions. However, from Fig.~\ref{starold} (upper-right panel) we find that the 0-12\% Au-Au/p-p ratio for {\em both} SS and AS is about six times that for p-p or peripheral Au-Au collisions---greatly exceeding what is implied by the ZYAM subtraction. 
A similar reconsideration of~\cite{staras} with reduced $p_t$ cuts might reveal that the ``away-side jet'' is still present, albeit with altered fragment (associated-particle) $p_t$ distribution.

\subsection{Dihadron correlations from PHENIX}

\begin{figure}[h]
  \includegraphics[width=3.3in,height=1.65in]{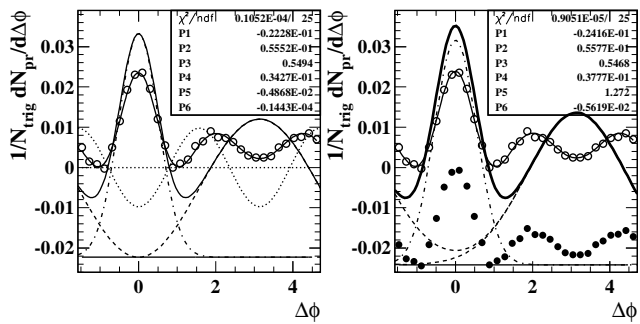}
\caption{\label{phenix}
Left: ZYAM-subtracted correlation data (open circles) with free fit (bold solid curve) of SS Gaussian (dash-dotted curve), AS dipole (dashed curve) and negative quadrupole (dotted curve) plus offset (solid line).
Right: The same data fitted with SS Gaussian (P2, P3), AS Gaussian (P4, P5) and quadrupole (P6), leading to equivalent results. The solid points are the open points translated by offset $P_1$.
 } %nf32  
 \end{figure}

Figure~\ref{phenix} (left panel) shows ZYAM-subtracted azimuth correlation data from Fig. 9 of~\cite{phenix} (points) corresponding to 2-3$\times$2-3 GeV/c $p_t$ cuts. The data are from 0-5\% central Au-Au collisions at $\sqrt{s_{NN}} = 200$ GeV. The bold solid curve is a free fit with offset (P1), SS Gaussian (amplitude P2, width P3) and away-side Fourier series (dipole P4, quadrupole P5, sextupole P6). The dash-dotted curve is the fitted SS Gaussian, the dashed curve is the fitted dipole, and the dotted curve is the fitted negative quadrupole. 

%%%%%%%%%%
%%%%%%%%%%%%

Figure~\ref{phenix} (right panel) shows a fit replacing the AS Fourier series with an AS Gaussian (amplitude P4, width P5) plus quadrupole (P6), which describes the data equally well. The SS peak parameters are the same in the two panels. In the left panel the Fourier series describing the AS peak is dominated by dipole and quadrupole. The sextupole $P_6$, which would be significant if an AS Gaussian were {\em required}, is negligible. The quadrupole is large and negative, reflecting mainly what was subtracted in the ZYAM procedure. The free fit recovers the correct AS jet structure and background offset. In the right panel the AS Gaussian contains the same information as the Fourier terms in the left panel. Gaussian width 1.3 is consistent with a pure dipole representation.

The quadrupole component of the unsubtracted data can be estimated for this case as well. The SS peak amplitude is $P_2 = 0.056$ and the width is $P_3 = 0.55$. The $2\,A_0\,v_2^2$ estimate is then $0.25\times 0.056 / 2 = 0.007$. The quadrupole amplitude after ZYAM subtraction is $-2\, P_5 = 0.01$, consistent with $v_2^2 \rightarrow 3\, v_2^2\{2\}/4$ and negligible net quadrupole component in the unsubtracted data (likely for 0-5\% central data).

Fig.~\ref{phenix} reveals that the increase in the SS peak yield over ZYAM subtraction is $(0.056\times 0.55) / (0.024\times 0.3) \sim 4$, and the increase in the AS yield is $(0.038\times 1.3) / (2\times 0.008\times 0.3) \sim 10$. The increase can be appreciated visually in the right panel by comparing the solid points to the bold solid curve. Again, jet yields are strongly suppressed by conventional ZYAM and $v_2$ subtraction.

%%%%%%%%
\section{Discussion}

I first discuss problematic aspects of conventional analysis and interpretation of triggered azimuth correlations prior to background subtraction. I then consider the consequences of ZYAM and v2 oversubtraction: underestimation of jet yields and distortion of jet correlations. Finally, I discuss contrasts with untriggered angular correlations and a proper fit model.

\subsection{The myth of the ``away-side'' jet}

In the context of so-called triggered jet analysis conventional language refers to an ``away-side'' (AS) jet~\cite{phenix,starprl,mach}. AS azimuth correlation structure is attributed to {\em single} jets which are complementary to ``high-$p_t$'' trigger particles (i.e., proxies for trigger partons).  The AS jet is said to broaden or, in the case of double-peaked structure, to manifest gluon bremsstrahlung radiation or Mach shocks~\cite{mach}. The description implies that two classes of jets exist: SS jets from the ``corona'' which are relatively intact (in-vacuum), and AS jets which suffer strong interactions with a dense medium (core), supporting the core-corona collision model. That language seems to represent a misunderstanding of jets in A-A collisions.

AS azimuth structure represents {\em inter}\,jet correlations, including not only {\em intra}\,jet structure (individual parton fragmentation) but parton-parton angular correlations as well, especially $k_t$ contributions (e.g., dijet acoplanarity). $k_t$ generation by initial-state parton multiple scattering in A-A collisions may dominate AS correlations there.

The SS jet peak on azimuth reveals  {\em intra}\,jet structure and represents {all} jets within an acceptance. Especially for untriggered angular correlations any modifications to jet structure (including Mach cones) should be manifested in the SS peak. Changes in AS correlations relative to the SS peak should first be attributed to the parton-parton relationship, dominated by $k_t$ production during initial-state N-N (parton-parton) scattering.

The role of $k_t$ (nucleon-intrinsic and from initial-state parton multiple scattering) even in p-p collisions is quite significant. The AS peak width in p-p collisions is typically 50\% larger than the SS peak, reflecting jet acoplanarity from the $k_t$ contribution to parton-parton transverse center-of-mass motion. The mean nucleon path length (as N-N encounters) is six in central Au-Au, implying up to six-fold increase in $k_t$ variance or 2.5-fold increase in AS peak width, which then accounts for typical AS widths $\sigma_\phi \sim 1.3-1.5$ in central Au-Au collisions. 

\subsection{Consequences of trigger/associated $p_t$ cuts}

Application of $p_t$ cuts to emulate jet reconstruction can strongly bias jet-related azimuth correlations, leading to misinterpretations. To interpret the consequences of $p_t$ cuts correctly a valid model of parton fragmentation (fragmentation functions or FFs in vacuum and medium) and parton-parton correlations ($k_t$ effects) is required.

For single-parton fragmentation (e.g., ``in-vacuum'' jets) in 200 GeV p-p collisions the following trends are observed~\cite{porter,2comp,fragevo}: 1) The peak (mode) of FFs occurs in 1-2 GeV/c for all parton energies from 3 GeV up to several hundred GeV~\cite{ffprd}. 2) p-p FFs are strongly suppressed below 1 GeV/c compared to jets in $e^+$-$e^-$ collisions. 3) About half the fragments from minimum-bias jets or ``minijets'' in p-p collisions, with most-probable parton energy = 3 GeV, appear below 1 GeV/c.

Some implications of those systematics are: 1) A typical trigger $p_t$ cut at 4 GeV/c eliminates {\em most} scattered partons, since $p_{t,parton} > p_{t,trigger}$, the mode of the parton spectrum is near 3 GeV and the spectrum is falling $\propto p_t^{-7.5}$. 2) FF suppression at smaller fragment momentum in p-p collisions already narrows jet angular correlations relative to in-vacuum jets from e-e collisions. 3) An associated $p_t$ cut at 2 GeV/c (e.g., ~\cite{staras}) eliminates most of the FF for any parton energy, further narrowing jet angular correlations (only the ``tip'' of the jet survives).

A trigger $p_t$ cut biases the underlying parton spectrum to larger energies and so could in principle be useful for systematic energy-loss studies. However, the trigger cut also biases the associated $k_t$ distribution, especially in more-central Au-Au collisions where accumulated $k_t$ can be large. Aside from jet (parton) acoplanarity ($k_t$ broadening of the AS azimuth peak) a unique consequence of the trigger cut is a $k_t$ bias in which the parton-pair $k_t$ sum is preferentially oriented nearly parallel to the trigger-parton momentum, thereby increasing the trigger-parton momentum and decreasing the partner momentum {\em in the laboratory frame}. Such $k_t$ bias can give an impression of large AS energy loss in a dense medium when the SS/AS momentum asymmetry is actually due to initial-state $k_t$.

In-medium modification of FFs in more-central Au-Au collisions can be described by alteration of the splitting process~\cite{bw}, wherein energy is not lost from a jet, is instead redistributed on $p_t$ to a degree depending on Au-Au centrality. More {\em jet-correlated} fragments appear at smaller $p_t$~\cite{fragevo}. The effect is observed in {untriggered} two-particle correlations, including a sharp transition at intermediate Au-Au centrality between unmodified p-p jets and strongly-modified but still intact jets in more-central Au-Au~\cite{daugherity,fragevo}. Modification of FFs in central Au-Au collisions can transport much of the jet below an associated-particle $p_t$ cut interval, giving the false impression that back-to-back {\em parton} pairs are suppressed, i.e., that one parton is ``thermalized'' in a dense medium. 

\subsection{Consequences of ZYAM and $\bf v_2$ oversubtraction}

This analysis demonstrates that for any case where there is significant SS and AS peak overlap the ZYAM assumption discards a significant fraction of the true jet yield. In more-central collisions up to 90\% of the jet yield may be discarded as a result of ZYAM plus $v_2^2$ oversubtraction, as in Fig.~\ref{phenix} (AS peak). Oversubtraction of $v_2^2$, which is guaranteed by conventional subtraction methods, further reduces the apparent jet yield and introduces strong distortions in AS correlations. The resulting distortions can be misinterpreted as evidence that a large fraction of scattered partons is thermalized in a dense medium, and that novel QCD phenomena result.

In~\cite{mach} the double-peaked AS structure induced in two-particle correlations is duplicated in a more-complex three-particle correlation analysis. The analysis is intended to demonstrate that relative to a trigger particle some pairs of AS particles are widely separated by a characteristic angle which might imply conical emission from a parton passing through a dense medium, as in Mach-cone shock waves~\cite{machth1,machth2}. 

It was found that the distance of AS ``Mach peaks'' from $\pi$ (``cone opening angle'') does not depend on A-A centrality or associated-particle $p_t$. The fixed opening angle has been interpreted to measure the speed of sound in the QCD medium~\cite{machth1,machth2}. Yet the properties of a true QCD medium ought to depend strongly on A-A centrality (e.g., compared to p-p collisions). The present analysis indicates that the AS double peak is an artifact having a fixed form for simple algebraic reasons, with no relation to collision dynamics or a QCD medium.

\subsection{Untriggered 2D angular autocorrelations}

Untriggered 2D angular autocorrelations~\cite{axialci,porter,ptscale,ptedep,daugherity} differ from conventional triggered dihadron correlation analysis in several ways: 1) Absent trigger/associated $p_t$ cuts the entire fragment yield from the entire minimum-bias parton spectrum is measured. 2) No {\em ad hoc} (ZYAM) offset is imposed; the offset is inferred from free fits to correlation structure. 3) Model fits impose only minimal {\em a priori} assumptions which are subsequently tested with data. 4) No external $v_2$ measurement is imposed; an independent quadrupole component is inferred directly from the model fit. 5) The SS peak is uniquely isolated in the 2D fit due to its $\eta$ dependence. 6) Remaining structure is described by model-independent azimuth Fourier amplitudes. The measured quadrupole amplitude serves as an upper limit to any nonjet quadrupole component.

Unsubtracted dihadron correlations for pairs within specific $p_t$ cut bins serve as the best measurement of {joint} distribution $v^2_2(p_{t1},p_{t2})$ simultaneously with corresponding jet structure. ZYAM subtraction disregards direct $v^2_2$ measurements, instead invokes products of $v_2(p_t)$ marginal values of questionable accuracy (large jet or ``nonflow'' contribution likely). Conventional $v_2(p_t)$ measurements for 0-5\% central Au-Au collisions imply $\sqrt{v_2(p_{t1})\, v_2(p_{t2})} \sim 0.06-0.07$, whereas untriggered 2D angular correlations return $v_2 \leq 0.01$. The difference in quadrupole amplitude $v_2^2$ is a factor 30-50.

The SS 2D Gaussian is tightly constrained by the 2D fit. Its amplitude is not available, as in a 1D fit, to offset distortions of the AS peak imposed by $v_2^2$ (quadrupole) oversubtraction. {\em No double-peaked AS structure has been observed in 2D angular autocorrelations}. 

 \subsection{Recommended procedures}

Whatever the subsequent analysis, unmodified dihadron pair distributions with absolute pair numbers per collision event should be published for all collision systems. A neutral presentation method for untriggered angular correlations normalizes the number of mixed pairs (reference) to the number of sibling (same-event) pairs~\cite{axialci,daugherity}. The integral of the net-pair distribution (number of ``correlated'' pairs) is made zero by construction to avoid suggesting that the net-pair number is known a priori. The true offset can be estimated subsequently via model fits. Mean event multiplicities for accepted particles should be included in the presentation.

The fit model should be as simple as possible (minimum parameter number) yet physically interpretable. Peaks can be distinguished from background by model fits as long as the peaks are resolved according to the Rayleigh criterion (sum of r.m.s. widths significantly less than peak separation $\pi$, which is typically satisfied for azimuth correlations). The Gaussian shape for the SS peak may be questioned in isolated cases (cf. Sec.~\ref{pythia}), but for most cases of relevance to RHIC data the 1D (on azimuth) or 2D Gaussian model should be adequate.

For higher $p_t$ cuts and/or p-p collisions the AS azimuth peak may be described by a Gaussian. For untriggered correlations and/or for more-central Au-Au collisions the AS peak width increases substantially, the higher Fourier components of the AS Gaussian decrease, and only a few terms of a Fourier series may be required. In the asymptotic limit (e.g., $\sigma_\phi \geq 1.3$) periodic AS peak structure simplifies to a pure dipole plus constant. 

In intermediate cases fits with both models (two Gaussians+quadrupole, Gaussian+dipole+quadrupole), including a sextupole in  the Fourier decomposition, provide optimal descriptions and estimates of systematic uncertainty. Fits should be performed over the unbroken $2\pi$ interval to respect and enforce periodicity of peak arrays. Analysis of SS or AS subintervals independently is inconsistent with discrete Fourier series and can be misleading.

%%%%%%%%%
 \section{Summary}

I have examined the validity of jet structure isolated from dihadron azimuth correlations by combinatoric background subtraction based on ZYAM (zero yield at minimum) and conventional $v_2(p_t)$ measurements. Simulations of ideal correlation data compared to published RHIC data demonstrate that in some cases ZYAM plus $v_2$ subtraction produces substantial distortions and underestimation of jet yields.

Combinatoric correlation analysis is a convenient method to access jet correlations in A-A collisions, but isolation of jet structure from the non-jet background presents technical challenges. The conventional background model includes a constant offset $B_0$ and an ``elliptic flow'' contribution nominally of the form $v_2^2(p_{t1},p_{t2})$. The ZYAM principle is invoked to determine $B_0$, but ZYAM has no basis in conventional distribution analysis. The ZYAM method is defended by arguing that the detailed shapes of jet correlations are not known a priori. Possible radiation effects and/or medium modifications to fragmentation could distort jet-related structure. ZYAM is then said to reduce systematic errors arising from uncertainties about jet peak shapes.

In this study I use simulations with known peak structures to demonstrate that the ZYAM procedure can produce large systematic errors. Systematic distortions may be much larger than any effects due to peak-shape uncertainties. Especially for more-central A-A collisions the ZYAM assumption leads to substantial underestimation of jet correlation amplitudes, interpreted in turn to imply large parton energy loss in a thermalized dense medium.

I demonstrate that for typical correlations from RHIC collisions simple peak models adequately describe jet correlation data. Intrajet correlations (same-side peak) are well-described by a Gaussian. In special cases a modified Gaussian could accommodate width fluctuations due to radiation effects. Interjet correlations (awayside peak, back-to-back jets) are well-described by a Gaussian or, if the AS width is large (typical), by an azimuth dipole component. That model permits accurate determination of the background offset value, in contrast to ZYAM.

Conventional measurements of $v_2(p_t)$ are susceptible to ``nonflow'' (jet correlations). In the conventional method of dihadron correlation analysis $v_2^2(p_{t1},p_{t2})$ is overestimated, especially in more-central A-A collisions. As a result of $v_2^2$ oversubtraction the same-side jet peak is further reduced, and the away-side peak acquires a minimum at its center. Away-side distortions have been interpreted as manifestations of Mach shocks arising from sound propagation in a thermalized dense medium.

As a result of ZYAM subtraction the impression may be formed that most partons are stopped in a thermalized dense medium (``opaque core''), that shock waves may be produced in the same medium as partons traverse the medium, giving rise to anomalous AS structure. In contrast untriggered angular correlations described by free model fits reveal that jet yields increase even faster than binary collisions in more-central Au-Au collisions. There is no isolated ``away-side'' jet. Away-side structure is undistorted and reflects jet-jet (parton-parton) correlations. Jets are modified in more-central collisions, but all partons survive as final-state correlation structure.

%%%%%%%%%%%%%%%%%%
This work was supported in part by the Office of Science of the U.S. DoE under grant DE-FG03-97ER41020.

%%%%%%%%%%%%%%%%%%%%%%%%%%%%

\end{document}